\pdfoutput=1
\documentclass[letterpaper]{JHEP3}

\let\ifpdf\relax
\usepackage{ifpdf}
\usepackage{comment}

\newcommand{\beq}{\begin{equation}}
\newcommand{\eeq}{\end{equation}}
\newcommand{\cO}{{\cal O}}
\newcommand{\cA}{{\cal A}}
\newcommand{\cAb}{{\overline{\cal A}}}

\newcommand{\cF}{{\cal F}}
\newcommand{\cFb}{{\overline{\cal F}}}
\newcommand{\cD}{{\cal D}}
\newcommand{\cDb}{{\overline{\cal D}}}
\newcommand{\cQ}{{\cal Q}}

\newcommand{\cU}{{\cal U}}

\newcommand{\cN}{{\cal N}}
\newcommand{\cUb}{{\overline{\cal U}}}

\newcommand{\Tr}{{\rm Tr\;}}

\newcommand{\phib}{{\overline{\phi}}}

\usepackage{graphicx}
\usepackage{amsmath}
\usepackage{amsfonts}
\usepackage{epsfig}
\usepackage{slashed}
\usepackage{epstopdf}
\usepackage{comment}

\def\bec{\begin{center}}
\def\eec{\end{center}}
\def\beq{\begin{equation}}
\def\eeq{\end{equation}}
\def\bea{\begin{eqnarray}}
\def\eea{\end{eqnarray}}

\title{Spontaneous supersymmetry breaking in two dimensional lattice super QCD}

\author{Simon Catterall and Aarti Veernala \\
Department of Physics, Syracuse University, Syracuse, NY13244, USA \\
}

\abstract{We report on a non-perturbative study of two dimensional $\cN=(2,2)$ super
QCD. Our lattice formulation retains a 
single exact supersymmetry at non-zero lattice spacing, and contains $N_f$ fermions in the fundamental representation of a
$U(N_c)$ gauge group. The lattice action we employ contains an additional Fayet-Iliopoulos term which is also invariant under
the exact lattice supersymmetry.
This work constitutes the first numerical study of this theory which serves as a toy model
for understanding some of the issues that are expected to arise in four dimensional super QCD. 
We present evidence that the exact supersymmetry breaks spontaneously when $N_f<N_c$ in agreement
with theoretical expectations.}

\begin{document}

\section{Introduction}

In recent years a new approach to the problem of putting supersymmetric theories on the lattice has been developed based on discretization of a topologically twisted
version of the continuum theory \cite{Sugino:2004qd,Catterall:2004np,Catterall:2005fd,physrep,D'Adda:2005zk,Damgaard:2008pa}.~\footnote{The same lattice theories can be obtained using orbifold methods and indeed supersymmetric lattice actions  for Yang-Mills theories were first constructed using
this technique \cite{Cohen:2003xe,Cohen:2003qw,Kaplan:2005ta,Damgaard} and the connection between twisting and orbifold methods forged in \cite{Unsal:2006qp}}
Initially the focus was on lattice actions that target pure super Yang-Mills theories in the continuum limit, in particular $\cN=4$  super Yang-Mills
\cite{latsusy-1,latsusy-2,Catterall:2012yq,Catterall:2011pd,Catterall:2013roa}. For alternative approaches to
numerical studies of $\cN=4$ Yang-Mills see refs.~\cite{Hanada:2013rga,Hanada:2010kt,Honda:2011qk,Ishiki:2009sg}.
However in \cite{Matsuura}~\cite{SuginoQuiver} these formulations
were extended to the case of  theories incorporating fermions transforming in the fundamental representation of the gauge group
and hence targeting super QCD. The starting point for these later lattice constructions is a 
continuum quiver
theory containing fields that transform as bifundamentals under a product gauge group $U(N_c)\times U(N_f)$. After discretization these  bifundamental fields connect two
separate lattices and, in the limit that the $U(N_f)$ gauge coupling is sent to zero, yield a super QCD theory
with a global $U(N_f)$ flavor symmetry. This construction is
described in detail in section 3.
The lattice action we have employed in this work includes an additional Fayet-Illopoulos term which, while invariant under the exact lattice supersymmetry, generates a potential for the scalar fields. It is straightforward
to show that this yields a non-zero vacuum expectation value for the auxiliary field (D term supersymmetry breaking) if $N_f<N_c$.  In section 4.
we show the results from numerical simulations of this theory which support this
conclusion; we measure a non-zero vacuum energy and show that a light state - the Goldstino- appears
in the spectrum of the theory if $N_f<N_c$. In contrast we show that vacuum energy is zero and this state is absent from
the spectrum 
when $N_f >N_c$ which is consistent with the prediction that the theory does not spontaneously break
supersymmetry in that case.

\section{The starting point: twisted $\cQ=8$ SYM in three dimensions}

We start from the continuum eight supercharge ($\cQ=8$) theory in three dimensions which is written
in terms of twisted fields which are completely antisymmetric tensors in spacetime under the twisted SO(3) group. The original two Dirac fermions
reappear in the twisted theory as the components of a K\"{a}hler-Dirac field
$\Psi=\left( \eta, \psi_{a}, \chi_{ab}, \theta_{abc} \right)$ where the indices $a,b,c=1 \ldots 3 $.
The bosonic sector of the twisted theory comprises  a complexified gauge field $\cA_a=A_a+iB_a$ containing the 
original gauge field $A_a$ and an additional vector field $B_a$. This additional field
contains the three scalars expected of the eight supercharge theory which, being vectors under the R symmetry,
transform as a vector field after twisting.
The corresponding action $S=S_{\rm exact}+S_{\rm closed}$ where
\bea
S_{\rm exact} &=& \frac{1}{g^2} \; \cQ \Lambda = \frac{1}{g^2} \; \cQ \int d^3x {\rm Tr} \left[ \chi_{ab}(x)\cF_{ab}(x) + \eta(x)\left[\cDb_{a},\cD_{a}\right] + \frac{1}{2}\eta(x)d(x) \right], 
\label{quiverActionQexact}\\
S_{\rm closed}&=& - \; \frac{1}{g^2} \int d^3x  {\rm Tr} \left[\theta_{abc}(x) \cDb_{[c}\chi_{ab]}(x) \right].
\label{quiverActionQclosed}
\eea 
Here all fields are in the adjoint representation of a $U(N)$ gauge group $X=\sum_{a=1}^{N^2} X_a T_a$ and we adopt an antihermitian basis for the
generators $T_a$. $\cD_{a}$ and $\cDb_{a}$ are the continuum covariant derivatives defined in terms of the complexified gauge fields as $\cD_{a} = \partial_{a} + \cA_{a}$ and $\cDb_{a} = \partial_{a} + \cAb_{a}$.
The action of the scalar supersymmetry on the fields is given by
\begin{eqnarray}
\cQ \cA_a &=&  \psi_a\nonumber\\
\cQ \cAb_a &=& 0\nonumber\\
\cQ \psi_a &=&  0 \nonumber \\
\cQ \chi_{ab} &=&  -\cFb_{ab}\nonumber\\
\cQ \eta &=&  d   \nonumber\\
\cQ \theta_{abc} &=& 0
\end{eqnarray} 
Notice that we have included an auxiliary field $d(x)$ that allows the algebra to be off-shell nilpotent $\cQ^2=0$. 
This feature
then guarantees that $S_{\rm exact}$ is supersymmetric. 
The equation of
motion for this auxiliary field is then
\beq
d(x)=\left[\cDb_{a},\cD_{a}\right]
\eeq
The $\cQ$-invariance of $S_{\rm closed}$ follows from the Bianchi identity\footnote{Note that it is also possible to write the 3d action completely in terms of an $\cQ$-exact form without a $\cQ$-closed term by employing
an additional auxiliary field $B_{abc}$} \\
\beq
\epsilon_{abc}\cDb_{c}\cFb_{ab} = 0.
\label{bianchi}
\eeq 
To discretize this theory we place all fields on the links of a lattice. This 3d lattice consists of the usual hypercubic vectors plus additional face and body links. In detail
these assignments are
\begin{center}
\begin{tabular}{c|c}\hline
continuum field & lattice link\\
$\cA_a(x)$ & $x\to x+\hat{a}$\\
$\cAb_a(x)$ & $x+\hat{a}\to x$\\
$\psi_a(x)$ & $x\to x+\hat{a}$\\
$\chi_{ab}$ & $x+\hat{a}+\hat{b} \to x$\\
$\eta(x)$ & $x\to x$\\
$d(x)$ & $x\to x$\\
$\theta_{abc}$ & $x\to x+\hat{a}+\hat{b}+\hat{c}$ \\
\hline
\end{tabular}
\end{center}
The lattice gauge field will be denoted $\cU_\mu(x)$ in the following discussion.
For the scalar fields $d(x)$, $\eta(x)$  the link degenerates to a single site. Notice that the orientation of a given fermion
link field is determined by the even/odd character of its corresponding continuum antisymmetric form. 
The link character of a field determines its transformation properties under lattice gauge transformations eg. $\cU_a(x)\to G(x)\cU_a(x)G^\dagger(x+\hat{a})$. 
To complete the construction of  the lattice action it is necessary to replace continuum covariant derivatives by appropriate gauged lattice difference operators. The necessary
prescription was described in \cite{physrep},~\cite{Matsuura},~\cite{twist2orb}. It is essentially determined by the simultaneous requirements that 
the lattice difference agree with the continuum derivative as the lattice spacing is sent to zero and that it yields expressions that transform
as the appropriate link field under lattice gauge transformations. The lattice difference operators acting on a field $f^{(\pm)_{a}}$, where $(\pm)$ corresponding to the orientation of the field\footnote{Note that $\psi_{a}(x)$ and $\theta_{abc}$(x) originate from lattice site x and are, thus, positively oriented. $\chi_{ab}(x)$, however, terminates at lattice site x and this therefore assigned a negative orientation.}, are given by:
\bea
\cD^{(+)}_{a}f^{(+)}_{b_{1},b_{2},...,b_{n}}(x) &=& \cU_{a}(x)f^{(+)}_{b_{1},b_{2},...,b_{n}}(x+\hat{a}) - f^{(+)}_{b_{1},b_{2},...,b_{n}}(x)\cU_{a}(x+\hat{b}) 
\label{diffop-1} \\
\cD^{(+)}_{a}f^{(-)}_{b_{1},b_{2},...,b_{n}}(x) &=& \cU_{a}(x+\hat{b})f^{(-)}_{b_{1},b_{2},...,b_{n}}(x+\hat{a}) - f^{(-)}_{b_{1},b_{2},...,b_{n}}(x)\cU_{a}(x) \\ 
\nonumber \\
\cDb^{(+)}_{a}f^{(+)}_{b_{1},b_{2},...,b_{n}}(x) &=& f^{(+)}_{b_{1},b_{2},...,b_{n}}(x+\hat{a})\cUb_{a}(x+\hat{b}) - \cUb_{a}(x)f^{(+)}_{b_{1},b_{2},...,b_{n}}(x) \\
\cDb^{(+)}_{a}f^{(-)}_{b_{1},b_{2},...,b_{n}}(x) &=& f^{(-)}_{b_{1},b_{2},...,b_{n}}(x+\hat{a})\cUb_{a}(x) - \cU_{a}(x+\hat{b})f^{(-)}_{b_{1},b_{2},...,b_{n}}(x) 
\label{diffop-2} \\
\nonumber \\
\cD^{(-)}_{a}f^{(\pm)}_{b_{1},b_{2},...,b_{n}}(x) &=& \cD^{(\pm)}f^{(\pm)}_{b_{1},b_{2},...,b_{n}}(x-\hat{a}) \\
\cDb^{(-)}_{a}f^{(\pm)}_{b_{1},b_{2},...,b_{n}}(x) &=&= \cDb^{(\pm)}f^{(\pm)}_{b_{1},b_{2},...,b_{n}}(x-\hat{a}),
\label{diffOps}
\eea where $\hat{b}=\sum_{i=1}^{n}\hat{b}_{i}$ in equations (\ref{diffop-1}) to (\ref{diffop-2}). 
For example the continuum derivative $D_a\psi_b$ becomes
\beq
\cD^{(+)}_a\psi_b(x)=\cU_a(x)\psi_b(x+\hat{a})-\psi_b(x)\cU_a(x+\hat{b})\eeq
This prescription yields a set of link paths which, when contracted with the link field $\chi_{ab}(x)$, yields a closed loop whose trace is gauge invariant:
\beq
{\rm Tr}\;\left[\chi_{ab}(x)\left(\cU_a(x)\psi_b(x+\hat{a})-\psi_b(x)\cU_a(x+\hat{b})\right)\right]\label{ex}\eeq
It has the
correct naive continuum limit provided that (in some suitable gauge) we can expand $\cU_a(x)=I_{N}+\cA_a(x)$. 
The field strength on the lattice, $\cF_{ab}(x)$, is defined using the forward difference operator as:
\beq
\cF_{ab}(x) = \cD^{(+)}_{a}U_{b}(x).
\label{Fab}
\eeq 
In lattice QCD the
unit matrix arising in this expansion is automatic since the link fields take their values in the group. However the constraints
of exact lattice supersymmetry require that the  lattice gauge fields take their values, like the fermions,
in the algebra. In this case the unit matrix can
then be interpreted as arising from giving a vev to the trace mode of the original scalar fields $B_a$. This feature is {\it required} by lattice supersymmetry
but is only possible because we are
working with a complexified $U(N)$ theory - another indication of the tight connection between twisting and exact
supersymmetry. It also implies that the path integral defining the quantum theory will
use a flat measure  rather than the usual Haar measure employed in conventional lattice gauge theory. Such a prescription
would usually break lattice gauge invariance but again complexification comes to the rescue since the Jacobian resulting from
a gauge transformation of the $D\cU$ measure cancels against an equivalent one coming from $D\cUb$. \\
\\ We now show how to use this three dimensional lattice model to construct a two dimensional quiver theory while maintaining the exact lattice
supersymmetry.

\section{Two dimensional quivers from three dimensional lattice Yang-Mills}

\begin{figure}[b]
\begin{center}
\includegraphics[height=80mm]{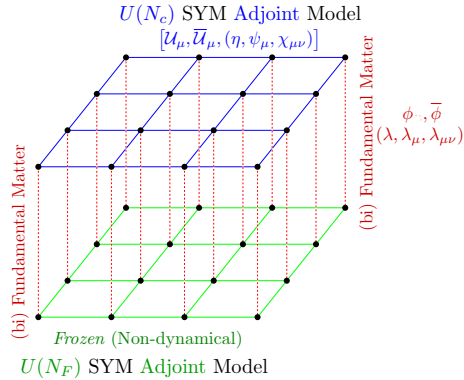}
\caption{3d quiver model}
\label{3dPic}
\end{center}
\end{figure}

Consider a lattice whose extent in the 3-direction comprises just two 2d slices.  Furthermore we shall assume free boundary conditions in the 3-direction
so that these two slices are connected by just a single set of links in the 3-direction - those running from
$x_3=0$ to $x_3=1$. Ignoring for the moment any fields that live on these latter links it is clear that the gauge group
can be chosen independently on these two slices. We choose a group $U(N_c)$ for the slice at $x_3=0$ and $U(N_f)$ at $x_3=1$ and will
henceforth refer to them as the $N_c$ and $N_f$ lattices. Denoting directions on the 2d slices by Greek indices
$\mu,\nu = 1,2 $ the fields living entirely on these lattices are given by
\bea
N_{c} \; \; \; &:& \; \; \; \Psi(x) = \left( \eta, \psi_{\mu}, \chi_{\mu\nu} \right), \; \; \; \cU_{\mu} = I_{N_{c}} + \cA_{\mu},\qquad d
\label{Nc-ferm} \\
N_{f} \; \; \; &:& \; \; \; \hat{\Psi} (\overline{x})= \left( \hat{\eta}, \hat{\psi}_{\mu}, \hat{\chi}_{\mu\nu} \right), \; \; \; \hat{\cU}_{\mu} = I_{N_{f}} + \hat{\cA},_{\mu}, \qquad \hat{d}
\label{Nf-ferm}
\eea 
In these expressions $x(\overline{x})$ denotes the coordinates on the $N_{c}(N_{f})$ lattice and $1_{N_{c}(N_{f})}$ denote the $N_{c}(N_{f}) \times N_{c}(N_{f})$ unit matrix respectively.
Now consider fields that live on the links between the $N_c$ and $N_f$ lattice. These must necessarily transform as bi-fundamentals under $U(N_c)\times U(N_f)$.
We have,
\bea
N_{c} \; \times \; N_{f}  \; \; \; &:& \; \; \;\Psi_{\text{bi-fund}}(x,\overline{x}) = \left( \psi_{3}, \chi_{\mu3}, \theta_{\mu\nu3} \right) = \left( \lambda, \lambda_{\mu}, \lambda_{\mu\nu}\right),\qquad \phi
\label{bifund}
\eea  The second equality in the above equation is 
a mere change of variables and corresponds to labeling fields according to their two dimensional character.
The complete field content of this model is summarized in the table below: \\
\begin{center}
\begin{tabular}{ c | c | c}
$N_{c}$-lattice & Bi-fundamental fields & $N_{f}$-lattice \\ 
$x$ & $(x,\overline{x})$ , $(\overline{x},x)$ & $\overline{x}$ \\ \hline
& & \\
$\cA_{\mu}(x)$ & $\phi(x,\overline{x})$   & $\hat{\cA_{\mu}}(\overline{x})$\\ 
$\eta(x)$ & $ \lambda(x,\overline{x})$  & $\hat{\eta}(\overline{x})$\\ 
$\psi_{\mu}(x)$ & $\lambda_{\mu}(\overline{x}+\mu,x)$ & $\hat{\psi}_{\mu}(\overline{x})$\\
$\chi_{\mu\nu}(x)$ &  $\lambda_{\mu\nu}(x,\overline{x}+\mu+\nu)$  & $\hat{\chi}_{\mu\nu}(\overline{x})$\\  
& & \\ \hline
\end{tabular} 
\end{center} 
Defining G(x) as a group element belonging to $U(N_{c})$ and H(x) to $U(N_{f})$ the lattice
gauge transformations for the bi-fundamental fields are as follows: 
\begin{eqnarray}
\phi(x) &\rightarrow& G(x)\phi(x)H^{\dagger}(\overline{x})\nonumber\\
\lambda(x) &\rightarrow& G(x)\lambda(x)H^\dagger(\overline{x})\nonumber\\
\lambda_{\mu}(x) &\rightarrow& H(\overline{x}+\mu)\lambda_\mu(x)G^{\dagger}(x)\nonumber\\
\lambda_{\mu \nu}(x) &\rightarrow &G(x)\lambda_{\mu \nu}(x)H^{\dagger}(\overline{x}+\mu+\nu)
\label{gaugetrans}
\end{eqnarray}
It is crucial to note that this generalization of the original lattice super Yang-Mills theory to a quiver model is completely consistent
with both the quiver gauge symmetries and the exact supersymmetry.
For example the 3d term given in eqn.~\ref{ex} yields a bi-fundamental term of the form
\beq  
{\rm Tr}\,\left[\lambda_{\mu}(x)\left(\cU_\mu(x)\lambda(x+\mu)-\lambda(x)\hat{\cU}_\mu(\overline{x})\right)\right]
\eeq which is invariant under the the generalized gauge transformations given in eqn.~\ref{gaugetrans}. 
Thus, the above construction lends us a consistent lattice quiver gauge theory containing both adjoint and bi-fundamental fields 
transforming under a product  $U(N_c)\times U(N_f)$ gauge group. Consider now setting the $U(N_f)$ gauge coupling to zero. This sets $\hat{\cU}_\mu=I_{N_{f}}$ up to gauge transformations and it
is then consistent to set all other fields on the $N_f$ lattice to zero. The original $U(N_f)$ gauge symmetry now becomes a global $U(N_f)$ flavor symmetry
which acts on a set of complex scalar fields $\phi$ transforming in the fundamental representation of the gauge group and their fermionic
superpartners $(\lambda,\lambda_\mu,\lambda_{\mu\nu})$. The situation is depicted in figure~\ref{3dPic}. 
At this point we have the freedom to add to the action one further supersymmetric and gauge invariant term - namely $r\sum_x \Tr d(x)=r \cQ \sum_x \Tr \eta$. This is a Fayet-Iliopoulos 
term. Its presence changes the equation of motion for the auxiliary field
\beq
d(x)=\cDb^{(-)}_\mu\cU_\mu(x)+\phi(x)\phib(x)-rI_{N_c}
\label{dFI} \eeq
with $I_{N_c}$ a $N_c\times N_c$ unit matrix. The SUSY transformations for the remaining adjoint and fundamental fields are:  
\begin{center}
\begin{tabular}{ c | c}
Adjoint Fields & Fundamental fields \\ \hline \\
$ \cQ \cA_{\mu} =  \psi_{\mu}$  & $ \cQ \phi = \lambda $ \\
$ \cQ \cAb_{\mu} = 0$ & $ \cQ \phib = 0 $ \\
$ \cQ \psi_{\mu} =  0 $ & $\cQ \lambda = 0$ \\
$ \cQ \chi_{\mu\nu} =  -\cFb_{\mu\nu}$ & $\cQ \lambda_{\mu} = -\cDb_{\mu} \phib$  \\
$ \cQ \eta = d$ & $\cQ \lambda_{\mu\nu} = 0 $ \\
&  \\
\end{tabular} 
\end{center}  After integration over $d$ the Fayet-Iliopoulos term
yields a scalar potential term which will play a crucial role in determining whether the system can undergo spontaneous supersymmetry
breaking.
The final action may be written as
\bea
S_{\rm adj} &=& \kappa \sum_{x} {\rm Tr} \left[ - \cFb_{\mu\nu}(x) \cF_{\mu \nu}(x) - \frac{1}{2} (\cDb^{(-)}_{\mu}\cU_{\mu})^{2}  - \eta(x) \cDb^{(-)}_{\mu}\psi^{\mu}(x) - \chi_{\mu \nu}(x)\cD^{(+)}_{[\mu}\psi_{\nu]}(x) \right], \nonumber \\
\label{action-adj} \\
S_{\rm fund} &=& \kappa \; \sum_{x} {\rm Tr} \left[-  \overline{\cD^{(+)}_{\mu}\phi(x)} \cD^{(+)}_{\mu}\phi(x) -  \frac{1}{2} \left[ \left( \phi(x)\phib(x) - \rm {rI} \right)^{2} \right] + \left[ \cDb_{\mu}^{(-)}\cU_{\mu}(x)\right] \left( \phi(x)\phib(x) - \rm {rI} \right)  \right]  \nonumber \\
 &-& \left[ \eta(x) \lambda(x) \phib(x) + \left \{ \lambda_{\mu}(x)\cD^{(+)}_{\mu}\lambda(x) -  \lambda_{\mu}(x)\psi_{\mu}(x) \phi(x+\mu) \right \} \right. \nonumber 
 \label{actionFundFerm-start} \\
 &-& \left. \left \{ \lambda_{\mu \nu}(x)\;\cDb_{\mu}^{(+)}\lambda_{\nu}(x) -  \lambda_{\mu \nu}(x)\phib(x+\mu+\nu)\chi_{\mu \nu}(x) \right \} \right],\nonumber \\
\label{actionFundFerm-end}
\eea 
In practice we have also included the following soft SUSY breaking mass term, $S_{\rm soft}$, in the adjoint action, $S_{\rm adj}$ in equation (\ref{action-adj}):
\beq
S_{\rm soft} = \mu^{2}\left[ \frac{1}{N_c} \Tr \left( \cUb_{\mu}\cU_{\mu} \right) - 1 \right]^{2}.
\label{actionSoft}
\eeq 
Such a term is necessary  to create a potential for the trace mode of the twisted scalar fields as we have discussed earlier. In principle we should extrapolate $\mu^2\to 0$ at the end of the calculation and so we have obtained all our results for
a range of $\mu^2$. In practice we observe that these soft breaking effects are rather small. \\
\\ Finally, the lattice coupling $\kappa$ appearing above is given by:    
\beq 
\kappa = \frac{N_{c}LT}{2\lambda A}. 
\eeq 	Here, $\lambda = g^{2}N_{c}$ is the dimensionful  `t~Hooft coupling, L and T are the numbers of points in each direction of
the 2d lattice and $A$ is a continuum area - the importance of
interactions in the theory being controlled by the dimensionless combination $\lambda A$. When we later discuss
our numerical results we refer to
this dimensionless combination as simply $\lambda$.

\section{Vacuum Structure and SUSY Breaking Scenarios}

Let us return to the equation of motion for the auxiliary field $d(x)$. If we sum the trace of this expression over all lattice sites and take
its expectation value we find
\beq
\langle \sum_x \Tr\,d(x) \rangle= \langle \sum_x \Tr\,\left(\phi(x)\phib(x)- rI_{N_c}\right) \rangle
\label{dvev}
\eeq
Since the lefthand side of this
expression is the expectation value of the $\cQ$-variation of some operator 
the question of whether supersymmetry breaks spontaneously or not is determined by whether the righthand side is
non-zero. Indeed after we integrate over the auxiliary field $d$ 
we find a scalar potential of the form
\beq
S_{\rm Dterm} = \sum_{x,f=1}^{N_{f}} \frac{\kappa}{2} \Tr \left( \phi^f(x)\phib^f(x) - \rm {rI_{N_c}} \right)^{2}, 
\label{Dterm}
\eeq 
Consider the
case where $N_f<N_c$.
Using $SU(N_c)$ transformations one can diagonalize the $N_c\times N_c$ matrix $\phi\phib$. In general it will have
$N_f$ non-zero  real, positive eigenvalues and $N_c-N_f$ zero eigenvalues.  This immediately implies that there
is no configuration of the fields $\phi$ where the potential is zero. Indeed the minimum of the potential will
have energy $r^2(N_c-N_f)$ and corresponds to a situation where $N_f$ scalars develop vacuum expectation values breaking the gauge group to $U(N_c-N_f)$. The situation when $N_f\ge N_c$ is qualitatively different;
now the rank of $\phi\phib$ is at least $N_c$ and a zero energy vacuum configuration is possible. In such a situation
$N_c$ scalars pick up vacuum expectation values and the gauge symmetry is completely broken. \\
\\ For the case when $N_f<N_c$ where $\cQ$-supersymmetry is expected to break  we would 
expect the spectrum of the theory to contain a massless fermion - the \emph{goldstino} \cite{Goldstino-theorem}.  To
see how this works in the twisted theory consider the vacuum energy
\beq
\langle 0| H |0\rangle \ne 0, 
\label{susy-breaking-1} 
\eeq which is equivalent to  $<\left \{ \cQ, \cO \right \}>\ne 0$ for some operator $\cO$. 
In the two dimensional twisted theory the relevant part of
the supersymmetry algebra is $\left \{ \cQ, \cQ_{\mu} \right \} = P_{\mu}$ \cite{latsusy2d} so that eqn.~\ref{susy-breaking-1} is
equivalent to
\beq
\langle 0| \left \{ \cQ, \cQ_{0} \right \} |0\rangle \ne 0,
\label{susy-breaking-3}
\eeq 
Note that the equation above involves both the scalar $\cQ$ and the 1-form supercharge $\cQ_{\mu}$. Corresponding
to these supercharges are a set of supercurrents, $ J$ and $J_\mu$ whose
form can be derived in the usual manner by varying the continuum twisted action under infinitesimal spacetime
dependent susy transformations. This yields gauge invariant supercurrents on the lattice of the  following form
\bea
J(x) &=& \sum_\mu \left[\psi_{\mu}(x)\cUb_{\mu}(x) \right] d(x) + ... ,
\label{J(x)} \\
J_0(x) &=& \eta(x)d(x) + ... ,
\label{Jprime(y)}
\eea 
and using the equations of motion, the auxiliary field d(x) can be replaced by
\beq 
d(x) = \sum_{\mu=1,2} \left[\cDb_{\mu}, \cD_{\mu} \right] + \left[ \phi(x)\phib(x) -rI_{N_c} \right] 
\eeq We therefore expect a possible Goldstino signal to manifest itself in the contribution of
a light state to the  two-point function:
\beq
C(t)=\langle 0| \cO(x) \cO^{\prime}(y) |0\rangle, 
\eeq where `t' corresponds to $(x^{0}-y^{0})$ and a suitable set of lattice interpolating operators are given by:
\beq
\cO(x) = {\rm Tr} \,\left[ \sum_\mu \psi_{\mu}(x)\cUb_{\mu}(x) \left( \phi(x)\phib(x) - {\rm rI_{N_{c}}} \right) \right]. 
\label{goldstino-2}
\eeq and 
\beq
\cO^{\prime}(y) ={\rm Tr} \,\left[ \eta(y)  \left(\phi(y)\phib(y) - {\rm rI_{N_{c}}} \right)\right].
\label{goldstino-3}
\eeq


\section{Numerical Results}
We employ a RHMC algorithm to simulate our system having first replaced all the twisted fermions in our model by corresponding pseudofermions - see for
example \cite{OOcode}~\cite{david}. The simulations are performed by imposing anti-periodic (thermal) boundary conditions on the fermions along one of the two space-time directions. This is done to avoid running into the fermion zero modes resulting from the scalar component of the twisted fermion, $\eta$. As discussed in \cite{2dsignprob}~\cite{Catterall:2014vga} this
has the added benefit of ameliorating the sign problem for these lattice theories. This breaks supersymmetry explicitly
by a term that vanishes as the lattice volume is
increased. \\
\\ In this section, we contrast results from simulations
with $N_f=2$, $N_c=3$ corresponding to the predicted
susy breaking scenario with results from simulations with $N_f=3$, $N_c=2$ - the susy preserving case. We ran our simulations for three different values of the `t~Hooft coupling, $\lambda=0.5,1.0$ and 1.5 and observed the same qualitative behavior for the different values of $\lambda$. The results presented in this section correspond to $\lambda=1.0$. The FI parameter, r, is a free parameter and is set to 1.0 for the rest of the discussion. \\
\\ As a first check, we compared the expectation value of the bosonic action with the theoretical value obtained using
a supersymmetric Ward identity
\beq
<\kappa S_{\rm boson}> = \left(\frac{3}{2}N_{c}^{2} + N_{c}N_{f}\right)V .
\label{bActionQuiver}
\eeq In appendix A. we show how to compute this value. Figure~\ref{baction}  shows a plot of the bosonic action for various values of the soft SUSY breaking coupling $\mu$. In principle we should take the limit $\mu\to 0$ although it should be clear from
the plot that the $\mu$ dependence is in fact rather weak. We have normalised the data to its value obtained by assuming supersymmetry
is unbroken. 
The red points at the bottom of the figure denote the SUSY preserving case  and
it can be observed that they agree with the theoretical prediction.  This is to 
be contrasted with the case when $N_f<N_c$  denoted by the blue points which shows a large
deviation from eqn.~\ref{bActionQuiver} and is the first sign that supersymmetry is spontaneously broken in this
case. \\
\begin{figure}
\begin{center}
\includegraphics[height=80mm]{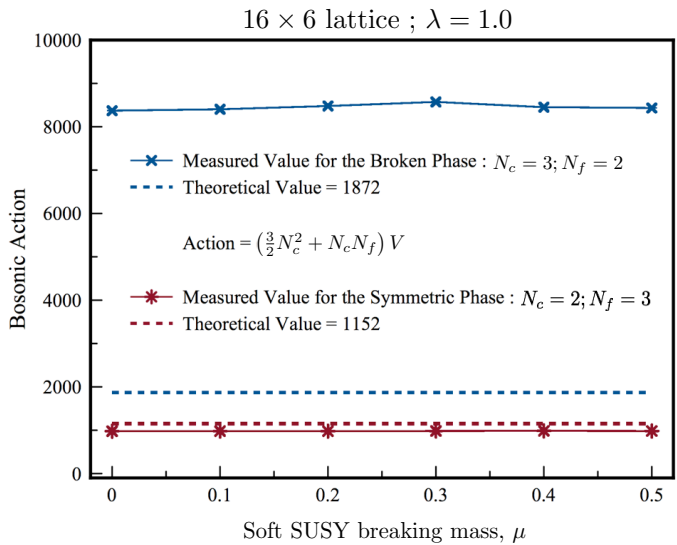}
\caption{Normalized bosonic action vs soft breaking coupling $\mu$ for $\lambda = 1.0$ for a 16x6 lattice}
\label{baction}
\end{center}
\end{figure}
\\ The spatial Polyakov lines shown in figure~\ref{polyS} also show a distinct difference 
between the $N_f<N_c$ and $N_f>N_c$ cases. The red lines where $|P|\approx 1$ correspond to the SUSY preserving case 
and are consistent with a deconfined or fully Higgsed phase.  Indeed the Polyakov line is
a topological operator and in a susy preserving phase should be coupling constant independent consistent
with what is seen. The blue line in the lower half of the plot corresponds to smaller
values which is qualitatively consistent with
the predicted partial Higgsing of the gauge field in the phase where supersymmetry is spontaneously broken. \\
\begin{figure}
\begin{center}
\includegraphics[height=80mm]{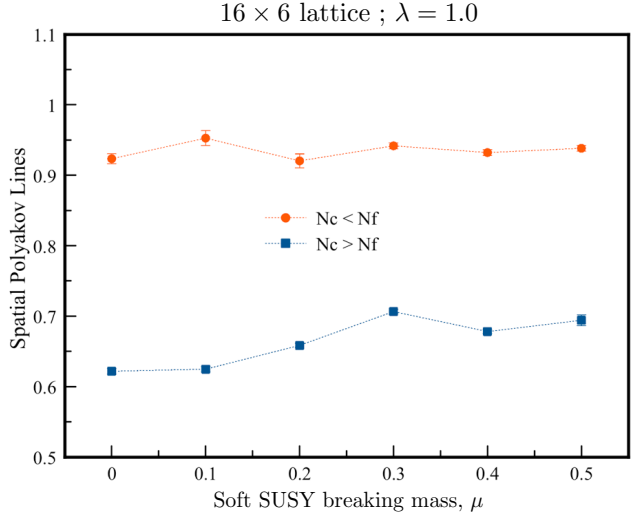}
\caption{Spatial Polyakov line vs $\mu$ for $\lambda=1.0$ on an 16x6 lattice}
\label{polyS}
\end{center}
\end{figure} 
\\ One of clearest signals of supersymmetry breaking can be obtained if one considers the equation of motion for the
auxiliary field eqn. \ref{Dterm}.  We expect the susy preserving case to obey
\beq
\frac{1}{N_{c}}  \Tr \left[ \phi(x)\phib(x) \right] = 1.
\label{Tr-ppd}
\eeq 
The red points, corresponding to ($N_{f} > N_{c}$) are consistent with this 
over a wide range of $\mu$. We attribute the small residual devaition
as $\mu\to 0$ to our use of
antiperiodic boundary conditions which inject explicit $\cQ$ susy breaking into the system. 
The simulations with $N_f<N_c$ (blue points) however show a clear signal for spontaneous supersymmetry breaking with the value of
this quantity deviating dramatically from its supersymmetric value even as $\mu\to 0$.  \\
\begin{figure}
\begin{center}
\includegraphics[height=80mm]{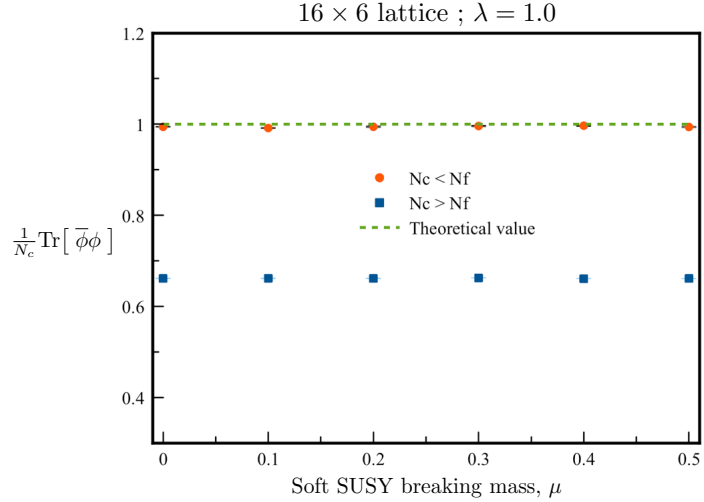}
\caption{$\frac{1}{N_c}{\rm Tr}\phi\phib$ vs $\mu$ for a 't Hooft coupling of $\lambda = 1.0$ on an 16x6 lattice}
\label{Tr-ppd-fig}
\end{center}
\end{figure} 
\begin{figure}
\begin{center}
\includegraphics[height=80mm]{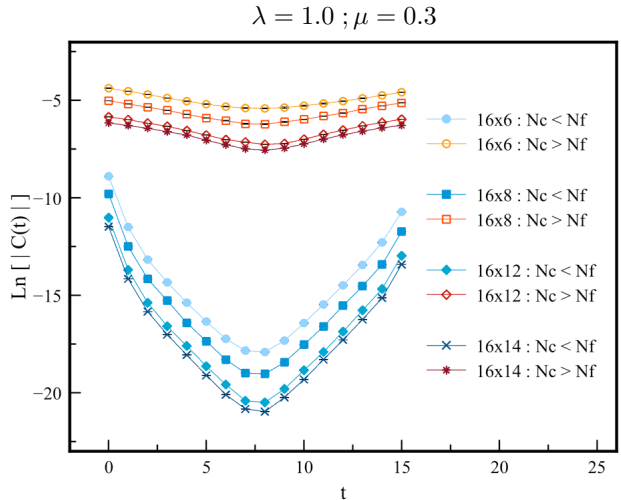}
\caption{Correlation function C(t)  for $\lambda = 1.0$ and $\mu = 0.3$ on various asymmetric lattices}
\label{effb}
\end{center}
\end{figure} 
\\ Finally we turn to our results for a would be Goldstino. We search for this by
computing  the following two point correlation function 
\beq
C(t) = \sum_{x,y}< O^{\prime}(y,t)O(x,0)> 
\label{C(t)}
\eeq where $O^{\prime}(y,t)$ and O(x,0) are fermionic operators given by:
\bea
O(x,0) &=& \psi_{\mu}(x,0)\cU_{\mu}(x,0)\left[\phi(x,0)\phib(x,0) - rI_{N_{c}} \right]
\label{Oprime} \\
O^{\prime}(y,t) &=& \eta(y,t) \left[\phi(y,t)\phib(y,t) - rI_{N_{c}} \right].
\label{O}
\eea Since it is computationally very cumbersome to evaluate the above correlation function for every lattice site x at the source we instead evaluate the correlator for every lattice site y for a few randomly chosen source points x. 
In figure \ref{effb} we show the
logarithm of this correlator as a function of temporal distance for a range of spatial lattice size, $L=6,8,12$ and 14. The anti-periodic boundary condition is applied along the temporal direction corresponding to T=16 and for both $N_f>N_c$ and $N_f<N_c$. The approximate linearity of
these curves is consistent with the correlator being
dominated by a single state in both cases. However when $N_f>N_c$ the
amplitude of this correlator is strongly suppressed  relative to the case where
$N_f<N_c$. Furthermore the effective mass  extracted from fits to this latter correlator (figure 6) falls as
the spatial lattice size (L) increases, consistent with  a vanishing mass in the large volume limit.  The lines in figure~6
show fits to $1/L$ - the smallest mass consistent with the boundary conditions - the dashed green line is
a fit constrained to go through the origin while the dotted red line allows the intercept to float.  This is just what
we would expect of a would be
Goldstino arising from spontaneous breaking of the exact $\cQ$-symmetry. 

\begin{figure}
\begin{center}
\includegraphics[height=80mm]{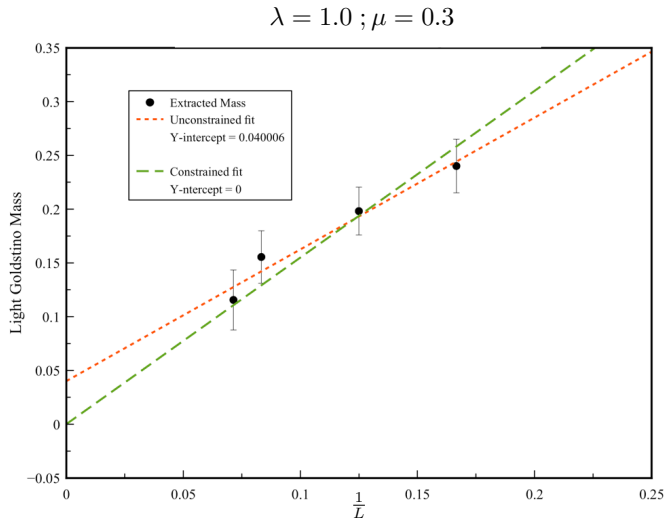}
\caption{Goldstino mass derived from fits $M_{eff}$ vs inverse transverse lattice size, $L^{-1}$}
\label{Meff}
\end{center}
\end{figure}


\section{Conclusions}

In this paper, we have reported on a  numerical study of super QCD in two
dimensions. The model in question possesses $\cN=(2,2)$ supersymmetry in the continuum limit
while our lattice formulation preserves a single exact supercharge for non zero lattice spacing. It is expected that
the single supersymmetry will be sufficient to ensure that full supersymmetry is regained without fine
tuning in the continuum limit. This
constitutes the first lattice study of a supersymmetric theory containing fields which transform
in both the fundamental and adjoint representations of the gauge group. Our lattice action
also contains a $\cQ$-exact Fayet-Iliopoulos
term which yields a potential for the scalar fields. The lattice theory possesses several
exact symmetries; $U(N_c)$ gauge invariance,
$\cQ$-supersymmetry and a global $U(N_f)$ flavor symmetry. \\
\\ It is expected that the system will
spontaneously break supersymmetry if $N_f<N_c$. The arguments that lead to this
conclusion depend on the inclusion of the Fayet-Iliopoulos 
term. Such a term is rather natural in our lattice model since the formulation requires
$U(N_c)$ gauge symmetry. Notice, though, that the free energy of the lattice model  does not naively
depend on the coupling $r$ as long as it is positive 
since the Fayet-Iliopoulos term is $\cQ$-exact.\footnote{In contrast for $r<0$ we
would expect supersymmetry breaking for any value of $N_f/N_c$. Thus one expects a phase
transition in the $N_f\ge N_c$ theory at $r=0$. }
Our numerical work is fully consistent with this picture; we have examined several supersymmetric Ward identities which
clearly distinguish between the $N_f<N_c$ and $N_f>N_c$ situations and we have observed a would be Goldstino state
in the former case. \\
\\ There are many directions for future work; inclusion of anti-fundamentals fields is straightforward
since it merely corresponds to including the bifundamental fields truncated from
the $N_f$-lattice. Observations of phase transitions in such models as the parameters
are varied can then potentially probe sigma models based on Calabi-Yau hypersurfaces  \cite{Witten:1993yc}. It is possible
that the $SU(N)$ theories could be studied by deforming the moduli space of the lattice
theory using ideas similar to those presented in \cite{new}. 
This would allow direct contact to be made to the continuum calculations of
Hori and Tong~\cite{HoriTong}. Finally the lattice  constructions discussed in this paper
generalize \cite{Joseph:2013jya} to three dimensional quiver theories leaving open the
possibility of studying 3D super QCD using lattice simulations.


\appendix
\section{Calculating the Bosonic Action}

Consider the partition function
\beq
Z=\int DX e^{-\kappa\left(\cQ\Lambda+S_c\right)}
\eeq
where $DX$ denotes the measure over all boson and fermion fields and $S_c$ the $\cQ$-closed
term. We start by rescaling the field $\theta_{abc}\to \kappa\theta_{abc}$ to remove the coupling
$\kappa$ from in front of the $\cQ$-closed term. This yields
\beq
Z=\kappa^{N_cN_fV}\int DX e^{-\kappa\cQ\Lambda-S_c}=\kappa^{N_cN_fV}Z^\prime \eeq
with $V$ the two dimensional volume. Notice that $N_cN_f$  is the number of fermions at each site resulting from the
3d $\theta$ field.
Differentiating with respect to $\kappa$ gives
\beq
-\frac{\partial\ln{Z}}{\partial\kappa}=-\frac{N_cN_fV}{\kappa}-\frac{\partial\ln{Z^\prime}}{\partial\kappa}\eeq
The last term in the righthand side being $\cQ$-exact would yield zero in the original theory containing
a $d$-field. However in the action we simulate this field is integrated out yielding instead a
contribution of $-\frac{1}{2\kappa} N_c^2V$
Putting these pieces together we find
\beq
\kappa<S_b>+\kappa<S_f>=-N_cN_fV-\frac{1}{2}N_c^2V\eeq
The expectation value of the fermionic action can be gotten by scaling arguments since the fermions
occur only quadratically in the action yielding
\beq
\kappa<S_f>=-\frac{4V}{2}\left(N_c^2+N_cN_f\right)\eeq
Collecting terms yields the final result quoted previously
\beq
\kappa<S_b>=V\left(\frac{3}{2}N_c^2+N_cN_f\right)\eeq

\acknowledgments SMC is supported in part by DOE grant
DE-SC0009998. SMC and AV would like to thank David Tong and David Schaich for useful discussions.
The simulations were carried out using USQCD resources at Fermilab.

\bibliographystyle{jhep}
\bibliography{quiverSQCDv15-jhep}

\end{document}